\documentclass[
    ,final            
  ]
  {aipproc}
\layoutstyle{6x9}

\newcommand\ion[2]{#1$\;${\small{#2}}\relax}%
\newcommand {\Lya}    {Ly$\alpha$}   
\newcommand {\Lyb}    {Ly$\beta$}    
\newcommand {\HI}     {\ion{H}{I}}   
\newcommand {\OVI}    {\ion{O}{VI}}   
\newcommand {\OVII}   {\ion{O}{VII}}
\newcommand {\OVIII}  {\ion{O}{VIII}}
\newcommand {\CIII}   {\ion{C}{III}}   
\newcommand {\NV}     {\ion{N}{V}}
\newcommand {\CIV}    {\ion{C}{IV}}
\newcommand {\SiIV}   {\ion{Si}{IV}}
\newcommand {\SiIII}  {\ion{Si}{III}}
\newcommand {\FeIII}  {\ion{Fe}{III}}
\newcommand {\NeVIII} {\ion{N}{VIII}}
\newcommand {\kms}    {km~s$^{-1}$}
\newcommand {\NHI}    {$N_{\rm HI}$}
\newcommand {\NOVI}   {$N_{\rm OVI}$}
\newcommand {\fOVI}   {$f_{\rm OVI}$}
\newcommand {\dndz}  {d{\cal N}\!\!/dz}

\newcommand {\Qhot}  {$Q_{\rm hot}$}
\newcommand {\OmegaWHIM} {$\Omega_{\rm WHIM}$}
\newcommand {\etal}  {et~al.}
\newcommand {\cd}    {cm$^{-2}$}
\newcommand {\lam}   {$\lambda$}

\begin{document}

\title[Low-$z$ Baryons]{Intergalactic Baryons in the Local Universe}

\classification{98.62.Ra}
\keywords      {Intergalactic matter; quasar absorption-line systems; O\,VI forest}

\author{Charles W. Danforth}{address={University of Colorado, CASA, 389-UCB, Boulder, CO 80309}}

\begin{abstract}
Simulations predict that shocks from large-scale structure formation and galactic winds have reduced the fraction of baryons in the warm, photoionized phase (the \Lya\ forest) from nearly 100\% in the early universe to less than 50\% today.  Some of the remaining baryons are predicted to lie in the warm-hot ionized medium (WHIM) phase at $T=10^5-10^7$~K, but the quantity remains a highly tunable parameter of the models.  Modern UV spectrographs have provided unprecedented access to both the \Lya\ forest and potential WHIM tracers at $z\sim0$, and several independent groups have constructed large catalogs of far-UV IGM absorbers along $\sim30$ AGN sight lines.  There is general agreement between the surveys that the warm, photoionized phase makes up $\sim30$\% of the baryon budget at $z\sim0$.  Another $\sim$10\% can be accounted for in collapsed structures (stars, galaxies, etc.).  However, interpretation of the $\sim100$ high-ion (O\,VI, etc) absorbers at $z<0.5$ is more controversial.  These species are readily created in the shocks expected to exist in the IGM, but they can also be created by photoionization and thus not represent WHIM material.  Given several pieces of observational evidence and theoretical expectations, I argue that {\it most} of the observed O\,VI absorbers represent shocked gas at $T\sim10^{5.5}$~K rather than photoionized gas at $T<10^{4.5}$~K, and they are consequently valid tracers of the WHIM phase.  Under this assumption, enriched gas at $T=10^{5-6}$~K can account for $\sim$10\% of the baryon budget at $z<0.5$, but this value may increase when bias and incompleteness are taken into account and help close the gap on the 50\% of the baryons still ``missing''.
\end{abstract}

\maketitle


\section{Introduction}

One of the major legacies of space-based UV instruments, in particular {\it HST} and {\it FUSE}, is the opening of the low-redshift intergalactic medium (IGM) to spectroscopic investigation.  Perversely, far more is known about the IGM at high-redshift than in the local universe, simply because many of the key diagnostic lines are rest-frame UV transitions only accessible to optical telescopes at $z\ge1.6$.  Indeed, before {\it HST}/FOS observations of low-redshift AGN, there was little interest in the low-$z$ IGM; the observed absorber frequency per unit redshift $\dndz$ of the \Lya\ forest lines in optical spectra drops as $(1+z)^{2.75}$ at redshifts of $z=2-4$ (Lu \etal\ 1991), thus predicting that \Lya\ systems at $z\sim0$ would be exceedingly rare.  However, FOS revealed a significant break in the steep power-law slope at $z\sim1.5$ (Weymann \etal\ 1998) due to the evolution of AGN ionizing radiation fields and the growth of large-scale structure.  The local IGM clearly provides astrophysical information of interest to cosmologists, and the sensitive UV spectrographs of the last decade have given us unprecedented opportunities.

Large-scale simulations (e.g., Dav\'e \etal\ 1999, 2001; Cen \& Ostriker 1999, 2006) predict that the baryon fraction $\Omega_{\rm Ly\alpha}$ in photoionized IGM (traced by \Lya\ forest lines) should drop from nearly unity at $z\sim4$ to a more modest fraction at local times as AGN radiation and large-scale structure shocks heat and ionize the IGM.  Concurrently, the mass in the warm-hot ionized medium (WHIM, $T=10^5-10^7$~K) is predicted to rise from nearly zero to a significant fraction of the total baryon budget at $z<0.5$.  The actual fractions in the different temperature and pressure phases depend sensitively on the methodology and parameters used in the simulations.  Therefore, establishing observational values for $\Omega_{\rm Ly\alpha}$ and \OmegaWHIM, as well as contributions from other phases at $z\sim0$, are of crucial importance to our understanding of everything from large-scale structure formation to galactic metal feedback to the distribution of AGN throughout cosmic evolution. 

Measuring the photoionized phase is relatively straightforward.  Despite the thinning of the \Lya\ forest at low $z$, \HI\ absorbers are still the strongest and most common in the IGM.  Several large surveys have established $\Omega_{\rm Ly\alpha}\approx 0.3\,\Omega_b$ at $z<0.5$ (e.g., Penton \etal\ 2000, 2004; Lehner \etal\ 2007; Danforth \& Shull 2008).  

Probing the WHIM phase is somewhat more complicated.  Hydrogen at these temperatures is almost entirely ionized, with no diagnostic spectral lines in any waveband.  Instead, we must rely on less direct tracers.  The first is to probe the vanishingly small neutral hydrogen fraction ($f_{\rm HI}<10^{-5}$ at $T\sim10^5$~K) through thermally-broadened \Lya\ lines in UV spectra (e.g., Richter \etal\ 2004, Lehner \etal\ 2007).  This method is fraught with many observational ambiguities and detection challenges (Danforth \etal\ 2009).

Tracing WHIM through X-ray transitions of common, highly-ionized metal ions such as \OVII\ and \OVIII\ (which peak in abundance at $T>10^6$~K) shows promise, but presents even more observational challenges.  Detections of a handful of \OVII\ and \OVIII\ absorbers in the IGM have been claimed toward several AGN (Nicastro \etal\ 2005, Fang \etal\ 2002, 2007), but none has yet been confirmed and much doubt exists as to their veracity (Kaastra \etal\ 2006, Rasmussen \etal\ 2007).  Current X-ray telescopes simply do not have the sensitivity and spectral resolution to detect the expected IGM absorbers.

\begin{table}[b]
  \begin{tabular}{lcrlc} 
    \hline 
      \tablehead{1}{c}{b}{O\,VI Survey}
     &\tablehead{1}{c}{b}{AGN}
     &\tablehead{1}{c}{b}{${\cal N}_{\rm OVI}$} 
     &\tablehead{1}{c}{b}{$\Delta z$}
     &\tablehead{1}{c}{b}{$\dndz$ ($>$30~m\AA)} \\
    \hline
     Danforth \& Shull (2005) & 31 & 40 & 2.2 & $17\pm3$             \\ 
     Danforth \& Shull (2008) & 28 & 83 & 5.2 & $15^{+3}_{-2}$       \\ 
     Tripp et al. (2008)      & 16 & 51 & 3.2 & $15.6^{+2.9}_{-2.4}$ \\ 
     Thom \& Chen (2008)      & 16 & 27 & 2.5 & $10.4\pm2.2$\tablenote{Blind survey, requiring {\it both} O\,VI lines for detection} \\ 
    \hline
  \end{tabular}
  \label{tab:a}
\end{table}

Far-UV transitions of highly-ionized metal ions (\OVI\ \lam\lam1032, 1038; \NV\ \lam\lam1238, 1242; \NeVIII\ \lam\lam770, 780, etc) offer the best current hope for WHIM measurement.  \OVI\ is the most promising of these with $\sim100$ confirmed IGM detections to date.  \OVI\ reaches a peak fractional abundance under collisional ionization equilibrium (CIE, Sutherland \& Dopita 1993) at $T\sim300,000$~K, and nothing short of soft X-ray photons ($h\nu>114$~eV) will produce it non-thermally.  However, as we will see, \OVI\ has a number of ambiguities as a WHIM tracer.

The idea of tracing WHIM with \OVI\ was pioneered in early work on individual AGN sight lines (Tripp \etal\ 2000, Savage \etal\ 2002, etc).  Each netted a few \OVI\ absorbers over a short redshift pathlength ($\Delta z<0.6$).  The largest survey to date of absorption lines in the low-$z$ IGM (including \OVI) was carried out by Danforth \& Shull (2008, hereafter DS08): a survey of 28 AGN sight lines with both {\it FUSE} and STIS/E140M data, following up the smaller Danforth \etal\ (2005, 2006) survey of 31 {\it FUSE} sight lines ($z_{\rm abs}<0.15$).  \Lya\ absorbers were identified interactively and checked for corresponding absorption in seven metal ion species (\OVI, \NV, \CIV, \CIII, \SiIV, \SiIII, and \FeIII).  The survey netted $\sim650$ \HI\ absorbers (many in \Lya, \Lyb, etc. enabling curve-of-growth analysis) over a redshift pathlength $\Delta z=5.2$, with 83 \OVI\ absorbers and smaller numbers of detections in other metal ions.  The \OVI\ detections showed $\dndz=15^{+3}_{-2}$ to a rest-frame equivalent width of $W_{1032}>30$ m\AA, or $\dndz=40^{+14}_{-8}$ to $W_{1032}>10$ m\AA.

Two other ambitious IGM \OVI\ surveys were published concurrently with DS08 (see Table~1), featuring many of the same AGN sight lines.  Tripp \etal\ (2008) searched 16 STIS/E140M datasets blindly for \OVI\ and surveyed \HI\ systems for \OVI\ absorption, resulting in 51 \OVI\ detections.  Despite differences in redshift pathlength definition and other systematic effects, they found surprisingly compatible results: $\dndz=15.6^{+2.9}_{-2.4}$ for $W_{1032}>30$ m\AA.  Thom \& Chen (2008) performed a true blind \OVI\ survey of 16 sight lines and found 27 \OVI\ absorbers for a frequency\footnote{Thom \& Chen (2008) required a detection in {\it both} O\,VI lines.  Their $\dndz$ value is correspondingly lower than both DS08 and Tripp \etal\ (2008) and can be taken as a lower limit.} $\dndz=10.4\pm2.2$.

\section{WHIM Baryons from O\,VI Absorbers}

DS08 derive the statistical fraction of low-$z$ baryons in the WHIM phase, $\Omega_{\rm WHIM}$, relative to the critical density $\rho_{\rm cr}$, by using \OVI\ absorbers as a proxy for WHIM baryons:
\begin{equation}    
  \Omega_{\rm WHIM}=\left(\frac{\mu m_{\rm H}\,H_0}{c\,\rho_{\rm cr}}\right) \frac{Q_{\rm hot}}{(O/H)_\odot\,(Z/Z_\odot)\,f_{\rm OVI}} ~\int_{N_{\rm min}}^{~\infty} \Big(\frac{d{\cal N}}{dz}\Big)\, \langle N_{\rm OVI} \rangle \, dN_{\rm OVI} \; ,   
\end{equation}
where \fOVI\ is the fraction of total oxygen ions in the five-times-ionized state (assumed to be $f_{\rm OVI}=0.22$, the CIE peak value) and $Z/Z_\odot$ is the typical metallicity of the IGM, taken here as 10\% solar.  Since \OVI\ can be produced either in $v>200$ \kms\ shocks or by X-ray photoionization ($h\nu>114$~eV), I introduce the parameter \Qhot\ in Eq. 1; the fraction of the observed \OVI\ that is shock-heated and thus represents legitimate WHIM gas.  In the analysis of DS05 and DS08, I assumed $Q_{\rm hot}=1$, but recent examples of photoionized \OVI\ are claimed by other groups.  Undoubtedly, \OVI\ is not wholly generated by either thermal ($Q_{\rm hot}=1$) or photoionized ($Q_{\rm hot}=0$) means, but I present arguments below that most observed \OVI\ is thermal in nature and thus the $Q_{\rm hot}\approx1$ approximation is appropriate.

\subsection{Shocks are Predicted to Exist}

One consistent feature of cosmological simulations is that shocks are generated by material falling from voids onto filaments in the cosmic web.  Typical infall velocities at $z<0.5$ are predicted to be 100--500 \kms, strong enough to produce post-shock temperatures of $10^5-10^7$~K (Fig.~1)\footnote{\tt http://solo.colorado.edu/$\sim$skillman/www/Cool\_Core.html}.  If the shocked gas is sufficiently enriched, observable \OVI\ absorption should be produced.  Feedback from galactic winds is important to reproduce the observed distribution of weak \OVI\ absorbers (DS08) through additional shock heating and IGM enrichment (Cen \& Fang 2006; Cen \& Ostriker 2006).

\begin{figure}[t]
  \includegraphics[height=.275\textheight]{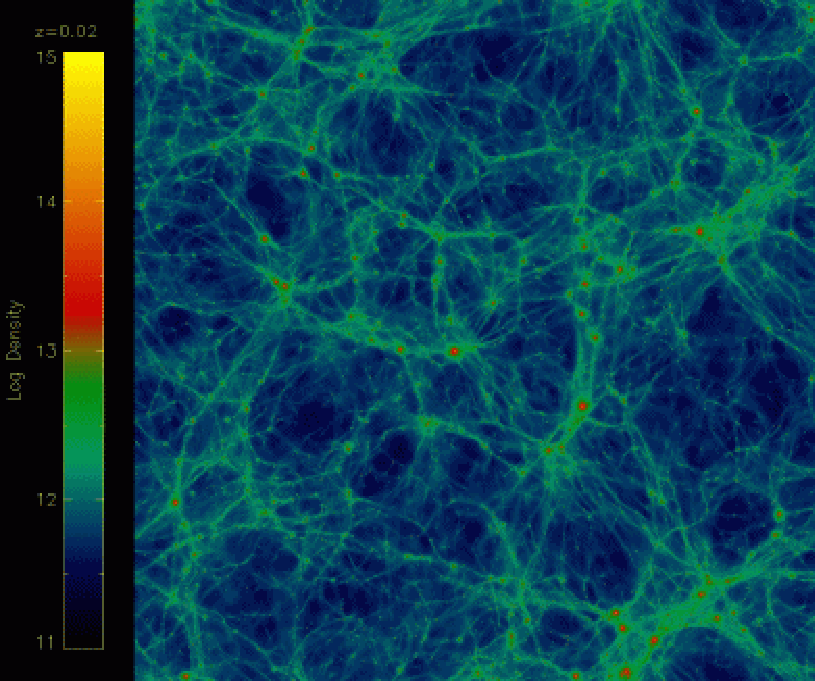}
  \includegraphics[height=.275\textheight]{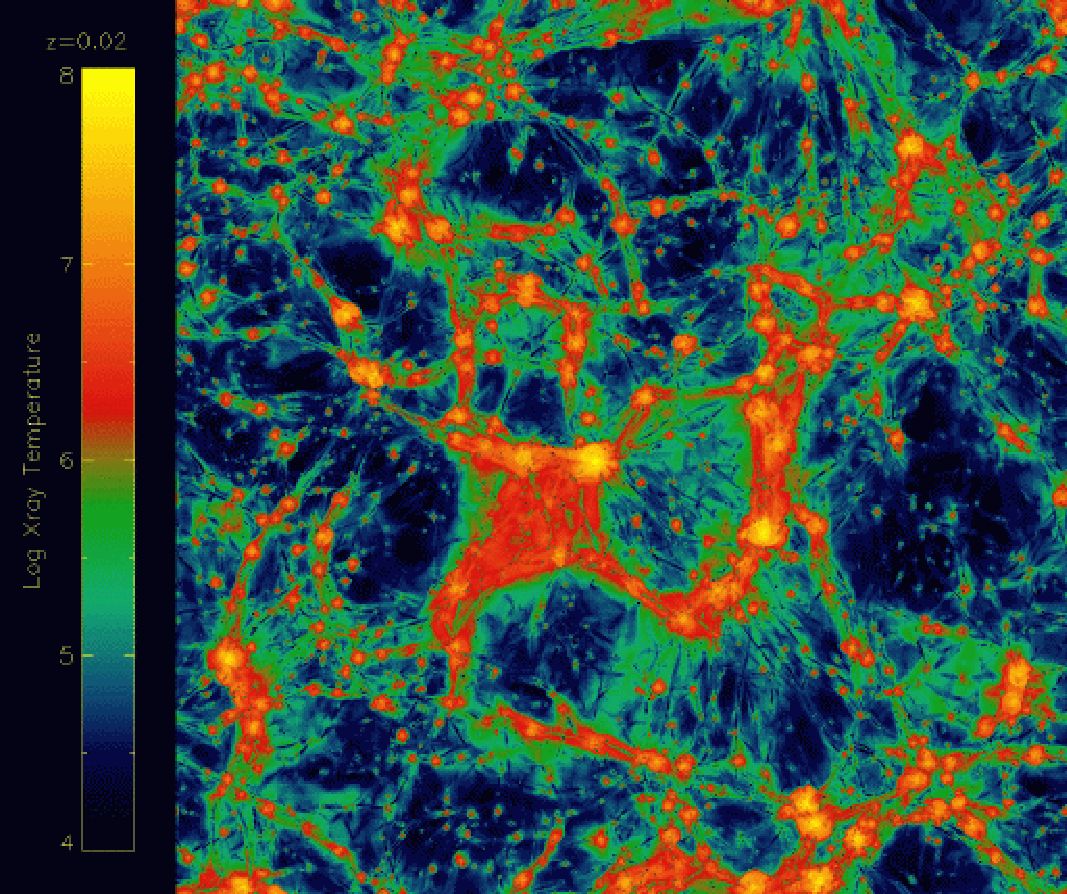}

  \caption{Shocks are predicted to exist in the IGM from infall of
  material onto filaments and galactic wind feedback.  Shown above are
  projections from a simulation by S. Skillman \etal (priv. comm.)
  showing density (left, [$M_\odot~Mpc^{-3}$]) and temperature (right,
  [K]).  Simulation size is 128 $h^{-1}$~Mpc.}
\end{figure}

\subsection{IGM Cooling Times are Long}

The IGM is diffuse enough that cooling times for hot gas will be very long (Collins, Shull, \& Giroux 2004):  
\begin{equation}
\tau_{cool}=\frac{3kT}{n\Lambda} \approx(60{\rm~Gyr})~T_6^{2.7}~Z_{0.1}^{-1}~\delta_{50}^{-1}.
\end{equation}
Cooling time is approximately proportional to $T_6^{2.7}$ (indexed to $T_6=10^6$ K) and inversely to metallicity, indexed to $0.1\,Z_\odot$ and overdensity $\delta_{50}\equiv \rho/\rho_{\rm cr}=50$ (equivalent to $n_H\sim10^{-5}~\rm cm^{-3}$ at $z\sim0$).  Shorter cooling times are possible if higher metallicity and density and lower post-shock temperatures are used, but it is difficult to get cooling in less than a Hubble time with realistic input parameters.

Since shocks are expected from simulations and cooling times are expected to be long, the observed \OVI\ gas could arise in the shock interface where gas is being collisionally heated to WHIM, rather than gas radiatively cooling back to ambient temperatures.  This explains both the velocity proximity of \OVI\ and lower ionization species and the narrow range of characteristic \NOVI\ values.  If the peak ionization species of postshock gas is \OVII\ ($T\sim10^6$~K), we would expect to see a much broader range of column densities and poor correlation with \HI.

\subsection{Poor Multiphase Correlation}

High ions studied in the far UV (e.g. \OVI, \NV) are usually seen within $\pm50$ \kms\ of the low-ionization species (\HI, \CIII, \SiIII, etc).  However, they exhibit poor correlation in column density, detection statistics, and $\dndz$ distribution.  DS05 and DS08 show that the correlation between \NOVI\ and \NHI\ is weak and that the different species cover vastly different ranges (1.5 and 3.5 orders of magnitude, respectively).  This is not true when comparing ions of similar species (e.g. \NV/\OVI\ or \HI/\SiIII) which are both well-correlated and with similar range.

High- and low-ionization species exhibit very different detection statistics.  Comparing \OVI, \CIII, and \SiIII\ absorbers (for which the best statistics exist at low-$z$), we see that high ions and low ions appear in the same systems only $\sim25-30$\% of the time.  More rigorously, these absorbers display a power-law behavior in absorber frequency per unit redshift per column density bin: $\dndz\propto N^{-\beta}$.  The power law slope $\beta$ for low ions and high ions are noticeably distinct: $\beta_{\rm HI}=1.73\pm0.04$, $\beta_{\rm SiIII}=1.8\pm0.1$, $\beta_{\rm OVI}=2.0\pm0.1$ for samples of 650, 53, and 83 absorbers, respectively.  This difference suggests a fundamental distinction between the high- and low-ion bearing gas.

\subsection{Incompatible Ionization Parameters}

\begin{figure}[b]
  \includegraphics[height=.23\textheight]{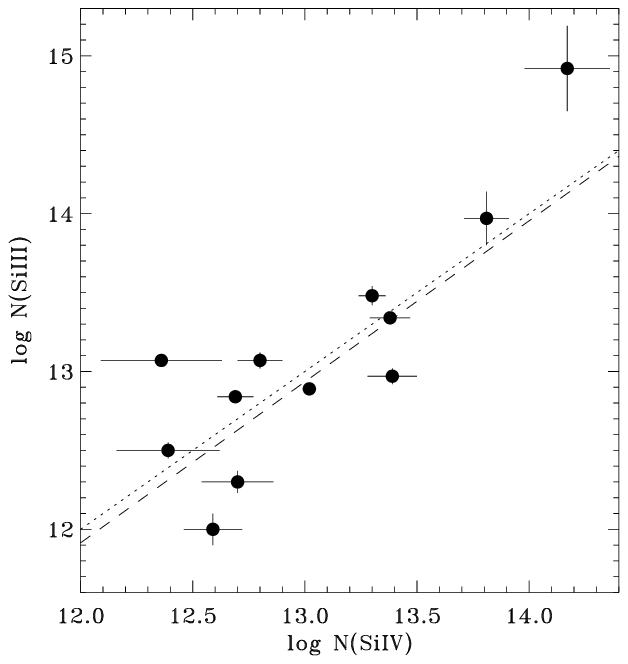}
  \includegraphics[height=.23\textheight]{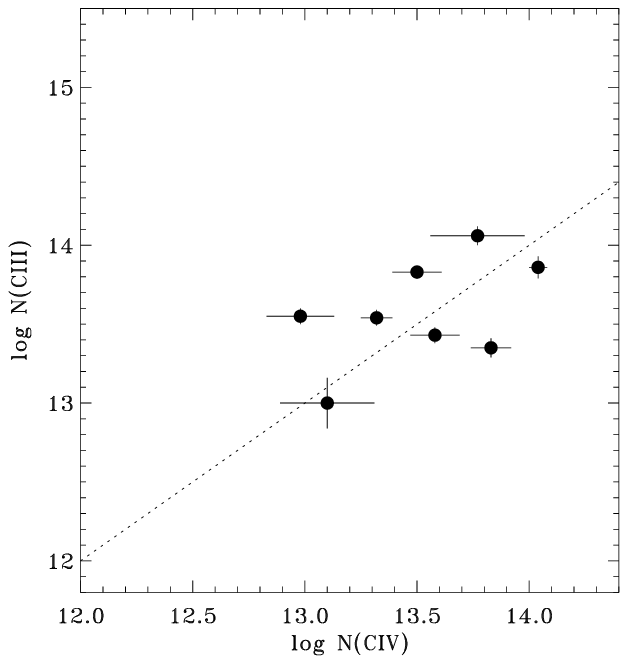}
  \includegraphics[height=.23\textheight]{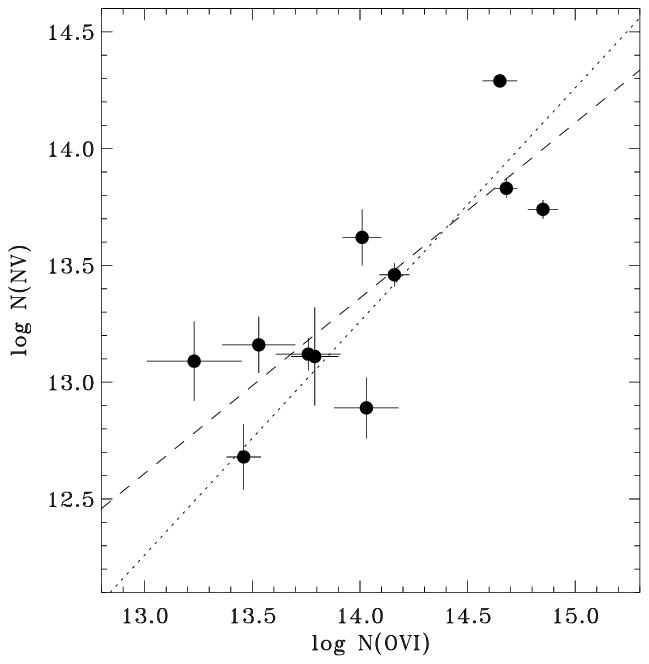}
  \caption{Column densities of ``adjacent'' ion stages Si\,III/IV (left), C\,III/IV (center) and N\,V/O\,VI (right) from DS08 are well-correlated, but require photoionization parameters that differ by a factor of $\sim10$ between the low and high ions.  Best fits (dashed) and linear relationships (dotted) are marked.}
\end{figure}

Another clue pointing to the distinct origin of high and low-ionization metal species comes from adjacent ion pairs such as C\,III/IV and Si\,III/IV which exhibit well-correlated column densities (Fig.~2).  Simple CLOUDY models give a required photoionization parameter $\log U\approx-1.5\pm0.5$, independent of IGM metallicity, density, and a reasonable range of ionizing spectral energy distribution (e.g., Telfer \etal\ 2002).  \NV\ and \OVI\ are different elements, but they correlate well in column density (Fig.~2).  Assuming a solar abundance ratio, the same CLOUDY models require either a photoionizing radiation field an order of magnitude stronger ($\log U=-0.4\pm0.5$) {\em or} high temperatures ($T=10^{5.4\pm0.1}$~K).

High and low metal ions are not usually detected in the same systems, so it should be no surprise that they require different ionization parameters.  The photoionization solution for \NV/\OVI\ ($\log U\sim-0.5$) requires a very diffuse IGM cloud ($n_H\sim10^{-6}~\rm cm^{-3}$) given what is known of the low-$z$ metagalactic ionizing field (e.g., Shull \etal\ 1999).  If a typical \OVI\ absorber has $N_{\rm OVI}=10^{14}~\rm cm^{-2}$, we can solve for the line-of-sight cloud dimension required to account for observed, photoionized \OVI:
\begin{equation}
\ell_{\rm abs}=\frac{N_{\rm OVI}}{n_H\,(O/H)_\odot\,(Z/Z_\odot)\,f_{\rm OVI}} \approx 3.5~{\rm Mpc}~\Big(\frac{N_{\rm OVI}}{10^{14}}\Big)\,\Big(\frac{0.1\,Z_\odot}{Z}\Big)\,\Big(\frac{10^{-6}}{n_H}\Big)\,\Big(\frac{0.2}{f_{\rm OVI}}\Big).
\end{equation}
An unvirialized cloud with these dimensions would display a line width of $\ell_{\rm abs}H_0\approx$~few$\times10^2$ \kms\ due to Hubble broadening alone, incompatible with the observed \OVI\ linewidths $\langle b_{\rm OVI}\rangle =25\pm15$ \kms.  It is possible to get the absorber dimensions down to sub-Mpc scales, but only by postulating less realistic parameters ($n_H$, $U$, $Z$): higher densities (and hence stronger ionizing fields) or higher metallicities.  On the other hand, collisionally ionized \OVI\ absorbers require only $\ell_{\rm abs}\approx200-350$~kpc to match the observations (e.g. Stocke \etal\ 2006, DS08).

\subsection{Arguments for Photoionized O\,VI}

Tripp \etal\ (2008) divide their 51 \OVI\ absorbers into ``simple'' and ``complex'' systems based on velocity alignment and profile shapes of \OVI\ and \HI\ components.  They argue that the simple absorbers represent gas in the same phase and use the line widths of both species to constrain the gas temperature.  In many cases, this temperature is less than $10^5$~K, suggesting that the simple \OVI\ systems do not represent WHIM baryons.

Component alignment is necessary, but not sufficient, to prove that the species are in the same phase.  One expects pre- and post-shock material to have a different velocity when looking through the plane of the shock.  However, sight lines parallel to the shock front will pass through both pre- and post-shock gas, but the velocity vector will be roughly in the plane of the sky.  Furthermore, supernova remnants and other expanding shock waves show considerably stronger emission/absorption along the edges (limb brightening with shock velocity transverse to the line of sight) than in the center where a large $\Delta v$ is expected.  The amplified column density in edge-on shocks means that we will see statistically more of them than face-on shocks with large velocity separations between pre- and post-shock gas.  Given the difficulties in generating \OVI\ by photoionization, this would seem to be the more logical conclusion.

Finally, Tripp \etal\ (2008) classify only 37\% of their \OVI\ systems as aligned, while the rest show velocity separation and are deemed multiphase.  Taking all the aligned systems as photoionized \OVI, via whatever assumptions are required to make it so, this still leaves $Q_{\rm hot}\approx0.63$.  If a fraction of the aligned systems are also multiphase, as I have argued above, \Qhot\ increases further and the $Q_{\rm hot}\approx 1$ approximation becomes valid.

\section{The Low-$z$ Baryon Census}

Thanks to concerted efforts by several groups, $\sim50$\% of the baryons in the local universe can now be accounted for as follows:

{\bf $\bullet$~~Stars, Galaxies, Virialized Mass: 7-10\%} e.g. Salucci \& Persic (1999)

{\bf $\bullet$~Warm Photoionized IGM (\Lya\ Forest): 29$\pm$4\%} Penton \etal\ (2000, 2004), Lehner \etal\ (2007), Danforth \& Shull (2008).


{\bf $\bullet$~Warm-Hot Ionized Medium (\OVI-WHIM): $\sim$10\%}.  \OVI\ absorbers serve as a proxy for metal-enriched gas at $T=10^5-10^6$~K (DS08), however this is probably a lower limit to the total WHIM fraction as discussed below.

{\bf $\bullet$~``Missing'': $\sim$50\%}  The balance of baryons may reside in one or more reservoirs as yet poorly explored observationally.  Possibilities include gas at $T>10^6$~K, metal-poor WHIM untraceable by metal line absorption, low column density absorbers (at any temperature phase), and intracluster gas.

\subsection{Biases, Uncertainties, \& Future Discovery Space}

Implicit in the current census of WHIM baryons are a number of uncertainties and biases.  Most of these serve to increase \OmegaWHIM\ and will undoubtedly account for {\em some} of the missing 50\% of the low-$z$ baryons once additional tracers are measured. 

{\bf $\bullet$  Photoionization correction ($Q_{\rm hot}<1$):}  As discussed above, I believe that the large majority of \OVI\ absorbers observed arise in shocked gas ($Q_{\rm hot}\approx1$).  However, some photoionized \OVI\ absorbers may exist.  Another possibility is that of a non-equilibrium state in which the cooling rate is faster than the recombination rate and the \OVI\ is ``frozen in'' to a high ionization state despite being in cool gas.  Both cases would result in non-WHIM \OVI\ ($Q_{\rm hot}<1$) and reduce \OmegaWHIM.

{\bf $\bullet$  (Z f$_{\rm ion}$) uncertainty:}  \OmegaWHIM\ scales inversely with metallicity $Z$ and ion fraction \fOVI\ (Eq.~1).  The assumptions of DS08 ($Z=0.1\,Z_\odot$, $f_{\rm OVI}=0.22$) are both likely upper limits for these parameters and thus \OmegaWHIM\ will increase when lower-metallicity systems and non-peak temperature conditions are considered.

{\bf $\bullet$  \Lya\ bias:}  DS08 measure \OVI\ absorption only at the redshifts of \Lya\ ``signposts''.  Thus, the population of high-ion absorbers without \Lya\ is unaccounted for in the present census.  Both Tripp \etal\ and Thom \& Chen perform ``blind'' surveys.  Tripp \etal\ find several ``naked'' \OVI\ systems, but all are plausibly intrinsic to the AGN rather than intervening IGM.  Thom \& Chen find one out of 27 \OVI\ absorbers with no \Lya\ counterpart; a close look at these data shows \Lya\ at low significance.  I conclude that the \Lya\ bias is likely small and has only a few percent effect on \OmegaWHIM.

{\bf $\bullet$ Metallicity bias:}  Low-metallicity and metal-free WHIM absorbers are likely if large-scale structure formation shocks are a major source of IGM heating.  Void metallicity is seen (Stocke \etal\ 2007) to be at least an order of magnitude lower than the canonical 10\% measured in DS08.  Galactic feedback shocks may account for a large fraction of what we see in \OVI\ and other metals.  Tracing low metallicity WHIM will require broad \Lya\ lines, with all the observational difficulties they represent.

{\bf $\bullet$ Temperature bias:}  Each FUV WHIM tracer is sensitive only to a narrow range of temperatures, with \OVI\ only common over $T=10^5-10^6$~K.  At high $T$, the plummeting hydrogen neutral fraction and rapidly broadening line profile make broad \Lya\ lines sensitive to only the coolest WHIM.  Given the shape of the cooling curve, hotter WHIM is expected to be more common, but it remains unmeasured in the current version of the baryon census.

{\bf $\bullet$ Weak absorbers:}  Finally, the observed \NOVI\ distribution $\dndz\propto N_{\rm OVI}^{-2}$ means that equal mass is contributed from many weak absorbers as from few strong ones.  Limits in sensitivity mean that there may be large reservoirs of WHIM in weak absorbers: $N_{\rm OVI}<10^{13}$~\cd\ ($W_{1032}<12$~m\AA).  High S/N COS observations of weak \OVI\ absorbers at $z>0.1$ will help to quantify this uncertainty.


\begin{theacknowledgments}
I would like to thank Mike Shull, Todd Tripp, Blair Savage, Teresa Ross, Eric Burgh and John Stocke for helpful input and critical review of this work  .  Figure~1 is courtesy of Sam Skillman.  Travel funding was generously provided by NASA/GSFC.  Support comes from COS grant NNX08-AC14G at the University of Colorado.
\end{theacknowledgments}

\bibliographystyle{aipprocl} 

\end{document}